\documentclass[conference,a4paper]{IEEEtran}
\IEEEoverridecommandlockouts
\usepackage{amsmath,amssymb,amsfonts}
\usepackage{algorithmic}
\usepackage{caption}
\usepackage{subcaption}
\usepackage{graphicx}
\usepackage{textcomp}
\usepackage{gensymb}

\usepackage{dblfloatfix} 

\usepackage{xcolor}
\usepackage{siunitx}
\usepackage{glossaries}
\usepackage{url}
\newacronym{med}{MED}{mean euclidean distance}
\newacronym{awgn}{AWGN}{additive white gaussian noise}
\newacronym{snr}{SNR}{signal-to-noise ratio}
\newacronym{sir}{SIR}{signal-to-interference ratio}
\newacronym{cdf}{CDF}{cumulative distribution function}
\newacronym{ml}{ML}{machine learning}
\newacronym{dl}{DL}{deep learning}
\newacronym{ism}{ISM}{image-source model}
\newacronym{rir}{RIR}{room impulse response}
\newacronym{en}{EN}{energy neutral}
\newacronym{cnn}{CNN}{convolutional neural network}
\newacronym{tof}{ToF}{time of flight}
\newacronym{hphr}{HPHR}{high precision high reliability}
\newacronym{ips}{IPS}{indoor positioning system}
\newacronym{pso}{PSO}{particle swarm optimization}
\newacronym{ls}{LS}{least squares}
\newacronym{tdm}{TDM}{time-division multiplexing}
\newacronym{fdm}{FDM}{frequency-division multiplexing}
\newacronym{lpf}{LPF}{low-pass filter}
\newacronym{pdf}{PDF}{probability density function}
\newacronym{crlb}{CRLB}{Cram\'er-Rao lower bound}
\newacronym{ga}{GA}{genetic algorithm}
\newacronym{gwo}{GWO}{gray wolf optimization}
\newacronym{nlos}{NLoS}{non-line-of-sight}
\newacronym{si}{SI}{swarm intelligence}
\newacronym{peb}{PEB}{position error bound}
\newacronym{fim}{FIM}{Fisher information matrix}
\newacronym{dop}{DOP}{dilution of precision}
\newacronym{mse}{MSE}{mean-square error}
\newacronym{rmse}{RMSE}{root-mean-square error}
\usepackage[english]{babel}
\usepackage[backend=biber,style=ieee,dashed=false]{biblatex }
\AtBeginBibliography{\footnotesize}
\usepackage{tabularx}
\newcolumntype{Y}{>{\centering\arraybackslash}X} 
\newcolumntype{L}{>{\raggedright\arraybackslash}X}
\newcolumntype{R}{>{\raggedleft\arraybackslash}X}
\usepackage{cellspace}
\usepackage{multicol}
\usepackage{multirow}
\usepackage{tablefootnote}
\usepackage{pifont}

\usepackage{bm}

\usepackage{url}

\usepackage[super]{nth}

\makeatletter
				\let\old@ps@headings\ps@headings
				\let\old@ps@IEEEtitlepagestyle\ps@IEEEtitlepagestyle
				\def\confheader#1{%
				\def\ps@headings{%
				\old@ps@headings%
				\def\@oddhead{\strut\hfill#1\hfill\strut}%
				\def\@evenhead{\strut\hfill#1\hfill\strut}%
				}%
				\def\ps@IEEEtitlepagestyle{%
				\old@ps@IEEEtitlepagestyle%
				\def\@oddhead{\strut\hfill#1\hfill\strut}%
				\def\@evenhead{\strut\hfill#1\hfill\strut}%
				}%
				\ps@headings%
				}
				\makeatother
				\confheader{%
                2023 13th International Conference on Indoor Positioning and Indoor Navigation (IPIN)}

\begin{document}

\title{Anchor Layout Optimization for Ultrasonic Indoor Positioning Using Swarm Intelligence\\
}

\author{\IEEEauthorblockN{Daan Delabie\IEEEauthorrefmark{1}, Thomas Wilding\IEEEauthorrefmark{4}, Liesbet Van der Perre\IEEEauthorrefmark{1}, Lieven De Strycker\IEEEauthorrefmark{1}}
\IEEEauthorblockA{\IEEEauthorrefmark{1}KU Leuven, WaveCore, Department of Electrical Engineering (ESAT), KU Leuven Ghent, 9000 Ghent, Belgium \\
\IEEEauthorrefmark{4}Graz University of Technology, Austria \\
daan.delabie@kuleuven.be }}

\maketitle

\begin{abstract}
Indoor positioning applications are craving for ever higher precision and accuracy across the entire coverage zone. Optimal anchor placement and the deployment of multiple distributed anchor nodes could have a major impact in this regard. This paper 
examines the influences of these two difficult to approach hypotheses by means of a straightforward ultrasonic 3D indoor positioning system deployed in a real-life scenario via a geometric based simulation framework. To obtain an optimal anchor placement, a \gls{pso} algorithm is introduced and consequently performed for setups ranging from 4 to 10 anchors. In this way, besides the optimal anchor placement layout, the influence of deploying several distributed anchor nodes is investigated. In order to theoretically compare the optimization progress, a system model and \gls{crlb} are established and the results are quantified based on the simulation data. With limited anchors, the placement is crucial to obtain a \gls{hphr} \gls{ips}, while the addition of anchors, to a lesser extent, gives a supplementary improvement.
\end{abstract}


\section{Introduction}\label{sec:intro}
\Glspl{ips} are aiming at an ever higher accuracy and reliability over the entire space of interest to enable applications such as object tracking in industry, hospitals, retail shops and venues. To obtain \gls{hphr} systems it is important, in addition to using better algorithms, to also scrutinize the anchor placement layout to which often none or little attention is paid. However, a thoughtful anchor placement layout dedicated to a given space can significantly increase accuracy and reliability~\cite{GA_example}. 

Typically there is not one optimal anchor placement layout that optimizes \gls{hphr} properties for all target positions, but there are several possibilities that show similar results~\cite{CRLBfirst}. Optimal anchor placement is mostly conducted for area-based localization \cite{OnOptimalAnchorPlacement, OnOptimal2}. To the best of our knowledge, optimal 3D anchor placement layout exploration has so far not been carried out, except in~\cite{Geometrical_ipin}, focusing on a minimum \gls{dop} value for a certain number of beacons.
In addition, the deployment of more distributed anchor nodes can increase spatial diversity, creating more redundancy to overcome reverberation and \gls{nlos} effects \cite{GA_example}. It is clear that in complex rooms with many obstacles, additional anchor nodes are desired. Typically, the more anchor nodes the better the accuracy, which is proven by measurements \cite{practical_amount_drone, acoustic_self_loc} and simulations \cite{acoustic_self_loc, optimalAhmad, PSOanchorsExample}. Mostly \glspl{crlb} are determined followed by a quest for the most closely related possible configurations via brute force searching or in an incremental way \cite{CRLBfirst}. Other researchers use methods such as \glspl{ga} \cite{GA_example} or \gls{si} algorithms \cite{optimalAhmad, PSOanchorsExample} to search for a viable solution.

Both the anchor placement puzzle and the desired amount of deployable anchors are difficult questions to solve. In this work we propose an optimization method based on \gls{si}, namely particle swarm optimization (PSO), whereby an optimal anchor placement layout in a practical 3D space is suggested for different numbers of anchors. The elaborated optimization method uses the variance of the 3D Euclidean distance error as a metric for the cost function. We strive for an optimal scenario in which the positioning accuracy is distributed as uniformly as possible over the entire room. Depending on the application, the optimal scenario can be defined differently, resulting in another optimization metric, e.g. the P95 value. For the positioning itself, a room simulation model~\cite{SimulationPaper} taking into account reverberation and directivities is used in which ultrasonic chirps are exploited for \gls{tof} measurements. The accuracy for 270 different locations in the room is determined in order to get the positioning in the entire room as reliable and accurate as possible.


To provide a clear picture of the proposed methods, the ultrasonic \gls{ips} is first revealed in Section~\ref{sec:ultrasonic_ips}.
Section~\ref{sec:optimization_algos} discusses suitable optimization algorithms to showcase the optimal anchor placement layouts.
These layouts are determined for scenarios with a different number of anchors in Section~\ref{sec:results}, and are compared to the obtained \glspl{crlb} that are established in Section~\ref{sec:crlb}. Concluding remarks are given in Section~\ref{sec:conclusion}.

\section{Ultrasonic Indoor Positioning System}\label{sec:ultrasonic_ips}
The \gls{ips} considered in the anchor placement problem uses ultrasonic chirp signals to perform \gls{tof} ranging, as elaborated in previous work \cite{mdpi_bert}. 
The speakers, acting as anchor nodes located at known positions $\bm{a}_j = [x_j,y_j,z_j]^\mathsf{T}$, sequentially send a \SI{30}{ms} chirp signal ranging from \SI{25}{kHz} to \SI{45}{kHz}. 
The mobile node, located at position $\bm{p}=[x,y,z]^\mathsf{T}$, is equipped with a microphone and receives a portion of the chirp (\SI{1}{ms}) at a given synchronization interval. Depending on the distance between the microphone and corresponding speakers, a different part of the chirp will be received. Upon reception, a correlation, i.e., a pulse compression, is performed between the received chirp part and the transmitted chirp. 
Given the planned synchronization, a time measurement and thus a ranging estimate can be made on the basis of the peak value of the correlation function. 
To better determine this peak value, a \gls{lpf} is applied to the absolute value of the correlation function. 
Due to reverberation, it is possible that no unambiguously correct peak can be determined, which means that the estimate may give incorrect results. 
In addition to peak selection methods as described in~\cite{IPINBert}, we here investigate the hypothesis that more redundancy can also be created when multiple distributed anchors are deployed, allowing for more accurate position estimates. To validate the latter, we focus for the ranging estimate on the maximum \gls{lpf} value. 
After the ranging estimates between the anchors and mobile node, a \gls{ls} positioning algorithm is applied to estimate the 3D position.

In reality, there are limitations in terms of placing anchor nodes in a given room, which is the motivation of this work. To this end, this study investigates the (constrained) problem of finding an optimal placement of $M$ anchors for the purpose of accurate positioning of a mobile device.
The focus lies on achieving a high positioning accuracy over the whole coverage area, with the performance analyzed at the example of the `Techtile' testbed environment, as described in~\cite{techtile-acoustic}.
To smoothly test different optimization techniques, this environment was simulated with the help of a realistic simulator which models reverberation and takes the directivity of nodes into account \cite{SimulationPaper}. 
For practical reasons, it is assumed that the anchors (speakers) can only be attached to the two largest side walls and the ceiling of the \SI{8}{\meter} $\times$ \SI{4}{\meter} $\times$ \SI{2.4}{\meter} room. The installation further requires that an anchor node has a $3~\mathrm{cm}$ distance from a plane, while a mobile node is at least $5~\mathrm{cm}$ away from a room plane.

\section{Anchor Layout Optimization algorithms}
\label{sec:optimization_algos}

An anchor placement solution can be found after an extensive search in which an optimization parameter is evaluated to yield positioning performance as close as possible to the \gls{crlb}, as mentioned in Section \ref{sec:intro}. 
Besides extensive search algorithms, heuristic solutions, such as \gls{pso} could pave the way to an optimal solution wherein the craved computational resources remain limited compared to an extensive search. Other heuristic methods could possibly be applied, such as \glspl{ga} \cite{GA_example} or \gls{gwo} \cite{PSOanchorsExample}, yet in the scope of this work we focus on the \gls{pso} which features are specifically suited for the problem at hand.

\subsection{Extensive Search Algorithm}
An extensive search to find a global optimum is very effective since practically all possible placement combinations for anchor nodes are tested. 
Depending on the size of the room and the size of the mutual distance between the anchors to be tested, the number of possible combinations can quickly increase. 
Assume in total $N$ potential anchor positions at the desired surfaces in a room, comprising the search space. $M$ of these anchor positions will be used for the deployment of anchors and thus positioning. 
In order to test all the combinations, $\binom{N}{M}$ simulations need to be performed. 
For example, if a set of 100 anchors is created in which all combinations of 6 anchors are tested, then $11.9 \!\times\! 10^8$ simulations must be performed and evaluated. Moreover, we want to test this not only for a selection of 6 anchors, but for $N\in \{4,6,\dots\}$ anchors whereby a set of 100 anchor nodes to perform an extensive search itself remains limited to really find the best option. Omitting the number of combinations that all lie in the same plane and are therefore unable to provide good 3D positioning only reduces the total number of combinations to be tested by a fraction. The fact that this brute-force approach may have a very large search space and thus comes at a huge computational complexity and hence is not practical for all applications, urges other optimization methods to be developed.

\subsection{Particle Swarm Optimization}
Finding a global optimum often requires an extensive search and is thus difficult to obtain, a sub-optimal or local optimum solution derived from a heuristic solution is preferred to find solid anchor placement layout results. 
A best solution is usually out of the question, but there are several (sub-)optimal layouts where similar results can be obtained. 
The positioning accuracy will be substantially better compared to a random anchor placement, allowing to avoid time consuming extensive search algorithms. 
As a heuristic method, we propose to apply \gls{pso} due to its simplicity and fast convergence, in addition to the ability to handle a wide range of functions and constrained problems \cite{PSOoverview}. 

The \gls{pso} algorithm is inspired by the behaviour of birds or fish, which move in flocks or schools in order to avoid predators or find food. 
In this work, the \gls{pso} algorithm initializes a population, termed a swarm, of different candidate solutions/anchor placements of $M$ anchor locations, as particles in the search space. 
To converge to an optimal solution, particles that are part of a given swarm update their relative positions iteratively in order to minimize a given cost function \cite{PSOoverview}. 
The movement of particle $i$ is guided by their personal best position $\bm{p}_{\text{best},i}$ and global best position $\bm{g}_{\text{best},i}$ of the swarm. 
The personal best position of particle $i$ are the $M$ anchor positions, stacked as $\bm{x}_i = [\bm{a}_{1,i}^\mathsf{T},\dots,\bm{a}_{M,i}^\mathsf{T}]^\mathsf{T}$, that results in the lowest cost function value (i.e. Euclidean distance error variance) over the iterations given by iteration indices $t\in\{1,2, \dots\}$. 
To determine the global best position, in addition to the best outcome of the past iterations, the best outcome of all the particles is also considered. 
The update performed between iterations is summarized as
\begin{align}
    \bm{v}_i^{t+1} &= \omega \bm{v}_i^t + c_1 \bm{r}_1 (\bm{p}_{\text{best},i}^t - \bm{x}_i^t) + c_2 \bm{r}_2 (\bm{g}_{\text{best},i}^t- \bm{x}_i^t)
    \label{eq:pso1} \\
    \bm{x}_i^{t+1} &= \bm{x}_i^t + \bm{v}_i^{t+1}
    \label{eq:pso2}
\end{align}
%
where $\bm{v}_i = [\bm{v}_{i,1}^\mathsf{T},\dots,\bm{v}_{i,M}^\mathsf{T}]^\mathsf{T}$ represents the velocity vector of particle $i$ which depends on the inertia weight $\omega$ representing a balance between local exploitation and global exploration, determining the partial influence that the previous velocity should have on the current movement. 
The coefficients $c_1$ and $c_2$ are the cognitive and the social coefficients respectively, which are multiplied with the uniformly distributed random vectors $\bm{r}_1$ and $\bm{r}_2$ to maintain diversity in the swarm population~\cite{PSO2}.
The choice of hyperparameters $\omega$, $c_1$ and $c_2$ needs to be determined based on a search approach for the specific application~\cite{PSO2}. 
After the update in Eq.~\eqref{eq:pso2}, some positions need to be reinforced again to remain within the search space, i.e, in our case, on certain room planes. 
Projections onto the nearest plane that is an element of the search space solve this problem. 
This also allows anchors to jump room planes instead of staying on their initialized plane.

\section{CRLB for ToF-based Indoor Positioning}\label{sec:crlb}
The \gls{crlb} is used as a reference to compare the performance obtained from the \gls{pso} algorithm over iterations and for different numbers and positions of anchors. 
Based on~\cite{GaussianBert}, we assume that the range measurements obtained from the correlation function discussed in Section~\ref{sec:ultrasonic_ips} can be modeled as the true distances $d_j(\bm{p})$ between mobile node and speakers $j\in\{1,\dots,M\}$, corrupted with \gls{awgn} noise~\cite{GaussianBert}.
This yields a range measurement model according to 
\begin{align}\label{eq:measurement_model}
r_j = \| \bm{p} - \bm{a}_j \| + n_j = d_j(\bm{p}) + n_j
\end{align}
where $n_j$ is the range measurement noise and $\|\cdot\|$ denotes the Euclidean norm.
The measurements for different speakers are assumed to be statistically independent, with $n_j$ modeled as zero mean Gaussian noise \cite{GaussianBert} with variance $\sigma_{\mathrm{r},j}^2$. 

The \gls{crlb} gives a lower bound for the variance of any unbiased estimator. 
It is calculated using the inverse of the \gls{fim} $\bm{J}(\bm{p})$ as \cite{kay1993} 
\begin{align}\label{eq:CRLB_def}
  \mathbb{E} [ (\hat{\bm{p}} - \bm{p}) (\hat{\bm{p}} - \bm{p})^\mathsf{T} ] \succeq \bm{J}(\bm{p})^{-1}
\end{align}
with the entry of row $m$ and column $n$ of the \gls{fim} defined as the partial derivative of the log-likelihood function $\ln f(\bm{r}|\bm{p})$ of all range measurements $\bm{r} = [r_1, \dots, r_J]^\mathsf{T}$. 
\begin{align}\label{eq:fim}
  [\bm{J}(\bm{p})]_{mn} = -\mathbb{E}\Big[\frac{\partial^2 \ln f(\bm{r}|\bm{p})}{\partial[\bm{p}]_m\partial[\bm{p}]_n}\Big]
\end{align}
Assuming independent measurements to all anchors, the measurement likelihood function factorizes as $f(\bm{r}|\bm{p}) = \prod_{j=1}^J f(r_j|\bm{p})$, with the $j$th anchors likelihood $f(r_j|\bm{p})$ being Gaussian due to the model from \eqref{eq:measurement_model}.
The \gls{crlb} allows to obtain a reference for performance comparison of the \gls{pso} algorithm described in Section~\ref{sec:optimization_algos}. 
Following the derivation from \cite{GuvencCST2009} one obtains
\begin{align}
[\bm{J}(\bm{p})]_{mn} = \sum_{j=1}^J \frac{1}{\sigma_{\mathrm{r},j}^2}\frac{[\bm{p}-\bm{a}_j]_m}{\| \bm{p} - \bm{a}_j \|} \;\frac{[\bm{p} - \bm{a}_j]_n}{\| \bm{p} - \bm{a}_j \|}.
\end{align}
The \gls{peb} for a mobile node at position $\bm{p}$ is then defined as the trace of the position \gls{crlb}  
\begin{align}
    \mathrm{PEB}(\bm{p}) = \mathrm{tr}[\bm{J}(\bm{p})^{-1}].
\end{align}

\section{Simulation Results}\label{sec:results}
This section discusses the simulation results of two investigated scenarios: the optimal placement of $M=4$ anchors showing the performance over \gls{pso} iterations, and the performance comparison of optimal anchor placements for multiple distributed anchors exploring $M=\{4,6,8,10\}$. 
The \gls{crlb} outlined in Section~\ref{sec:crlb} is used as performance reference. 

\subsection{Optimal Anchor Placement}\label{sec:anchor_placement}
The impact of the anchor placement layout is examined by comparing a non-optimized and optimized scenario. 
In both cases, 4 anchors are used to perform 3D indoor positioning for 270 mobile node positions, which are evenly distributed over the room on a regular grid. 
The spacing along the $x$, $y$ and $z$ axes of the mobile node locations is \SI{98.8}{cm}, \SI{78.0}{cm} and \SI{57.5}{cm} respectively. 
The anchors are placed in the room as described in Section~\ref{sec:ultrasonic_ips} and take into account the stated limitations. 
The sampling frequency of the signals received by each mobile nodes' microphone is chosen to be \SI{192}{kHz} to meet realistic hardware specifications for audio sampling. 
For the microphones, an omni-directional directivity is assumed, while the speakers apply a cardioid pattern to exhibit a realistic model. As an additional assumption, the directivity vector of the speakers always points towards the centroid of the room. 
The \gls{snr} is defined for the closest mobile node to a specific anchor and is chosen as \SI{30}{dB}. 
An equivalent amount of noise is then added to the received signals of the other mobile nodes from that specific anchor. 
This results in different noise standard deviation values for the different anchors since the distances to the closest mobile nodes are similar but not the same. 
For the obtained optimal anchor placement discussed in this section, the noise standard deviation for the 4 different anchors varies between $\sigma_\text{n}=0.0016$ and $\sigma_\text{n}=0.0079$. 
Due to the fact that the positions of the anchors change over iterations, these noise values will also change.

To perform the search for optimum anchor placement in terms of a high positioning accuracy at all investigated mobile node positions, the \gls{pso} algorithm from Section~\ref{sec:optimization_algos} is applied. 
At each iteration $i$ each particle $\bm{x}_{i}$ contains all positions of the 4 anchors. 
The swarm, in this case, counts 15 particles. 
The initialization phase creates these randomly placed particles, taking into account the constraints and assumptions. 
In the considered application, anchors can only be installed on the two largest walls and the ceiling. 
The anchors are distributed based on surface ratios with more anchors placed on larger surfaces. 
In the case of 4 anchors, this means that both walls get 1 anchor, and the ceiling the remaining 2. In all cases, the directivity is pointing towards the room centroid. 
During the \gls{pso} iterations, the simulations are performed to obtain a positioning improvement based on a proposed optimization metric/cost function. 
The variance of the Euclidean distance error of all position estimates is used as a metric, since the smallest possible error is desired over the entire space. 
The smaller the variance, the steeper the slope of the resulting \gls{cdf} or generally the smaller the chance of large observed outliers. 
When updating the anchor positions between iterations, it may happen that the anchors are no longer placed within the search space (on the certain surfaces). 
This is again enforced through projections onto the nearest planes as discussed in Section~\ref{sec:optimization_algos}. 
Subsequently, the directivity vector is adjusted. 
After several iterations, the improvement per iteration will drop below an arbitrary threshold, meeting the stopping criteria
\setlength{\belowcaptionskip}{-10pt}
\begin{figure}[t]
\centering
    \centering
    \includegraphics[width=0.75\linewidth]{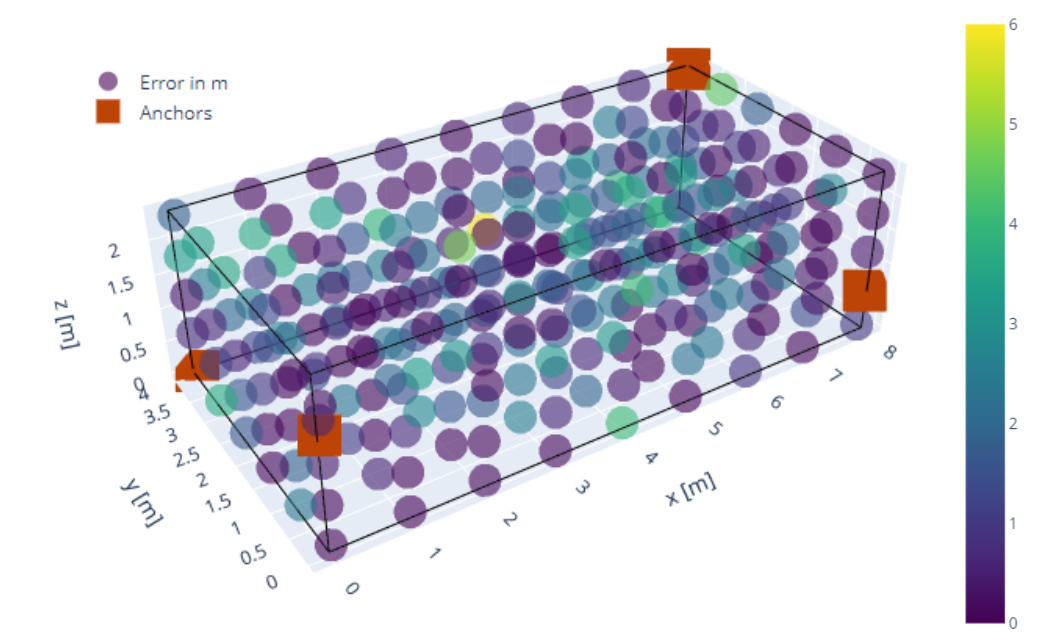}%
    \caption{Euclidean distance errors for the non-optimized 4~anchor placement.}%
    \label{fig:4anchscenarios1}
\end{figure}

\begin{figure}[t]
\centering
    \centering
    \includegraphics[width=0.75\linewidth]{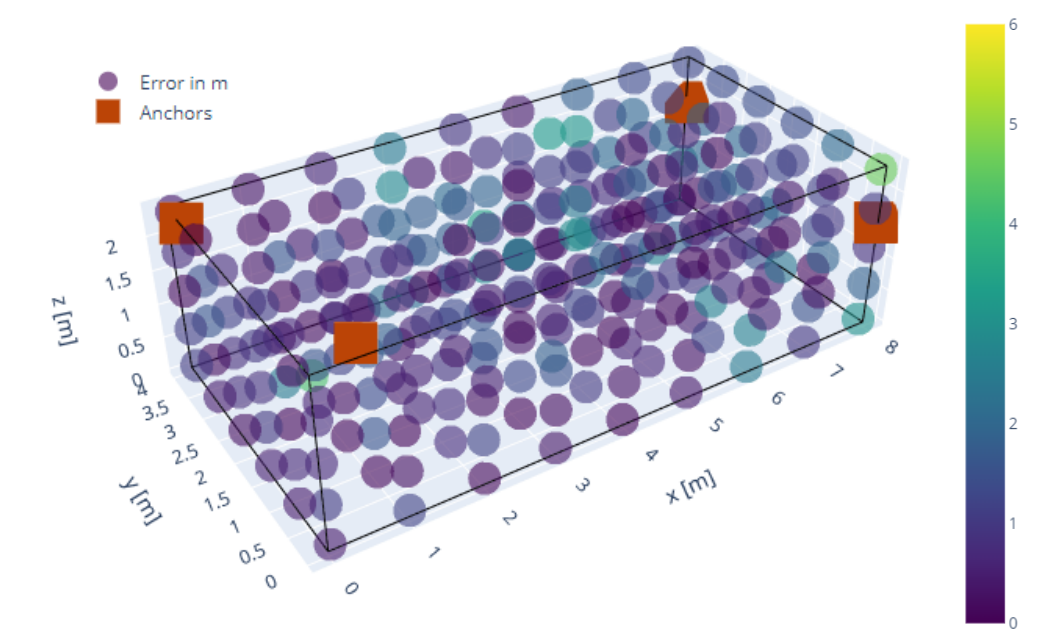}%
    \caption{Euclidean distance errors for the optimized 4 anchor placement.}%
    \label{fig:4anchscenarios2}
\end{figure}
\setlength{\belowcaptionskip}{0pt}

For the non-optimized scenario, the anchors are placed at the wall and ceiling corners of the room, as illustrated with the orange squares in Fig.~\ref{fig:4anchscenarios1}, so that when the neighboring anchors are fictitiously connected together with a line, the envelope curve encompasses the room. 
Intuitively, it seems that this should give better positioning accuracy results since the mobile nodes are within the envelope of the anchors.

The 3D Euclidean distance errors between the real and estimated positions and the anchor placements for the non-optimized and optimized scenarios are shown in Fig.~\ref{fig:4anchscenarios1} and Fig.~\ref{fig:4anchscenarios2}. 
The standard deviation values for the non-optimized and optimized scenarios are respectively \SI{1.362}{m} and \SI{0.811}{m}. It can be seen that in general the accuracy of almost all position estimates increased after optimization, at the expense of decreased accuracy for a few positions. 
Most notably, large outliers are eliminated. 
An optimal scenario for all positions, which strongly depends on the application and optimization parameter, is not possible but it can be aspired to be achieved. 

\begin{figure}[t]
\centering
    \centering
    \includegraphics[width=0.82\linewidth]{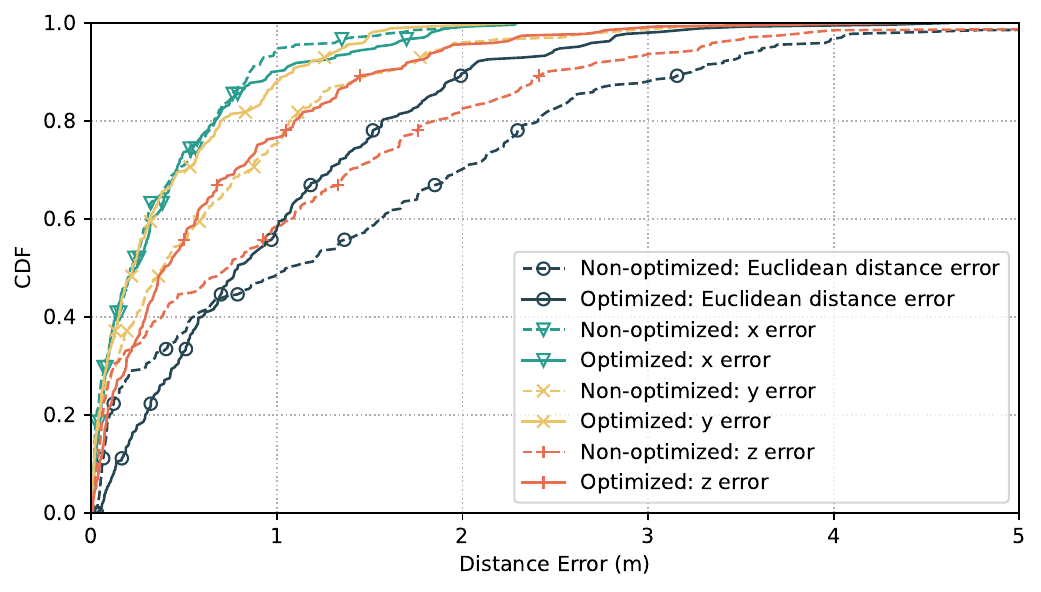}%
    \caption{Comparison between the Euclidean 3D distance, x, y and z error values obtained by the optimized and non-optimized \gls{ips}.}%
    \label{fig:4anchCDFcurves}
\end{figure}

\begin{figure}[t]
\centering
    \centering
    \includegraphics[width=0.75\linewidth]{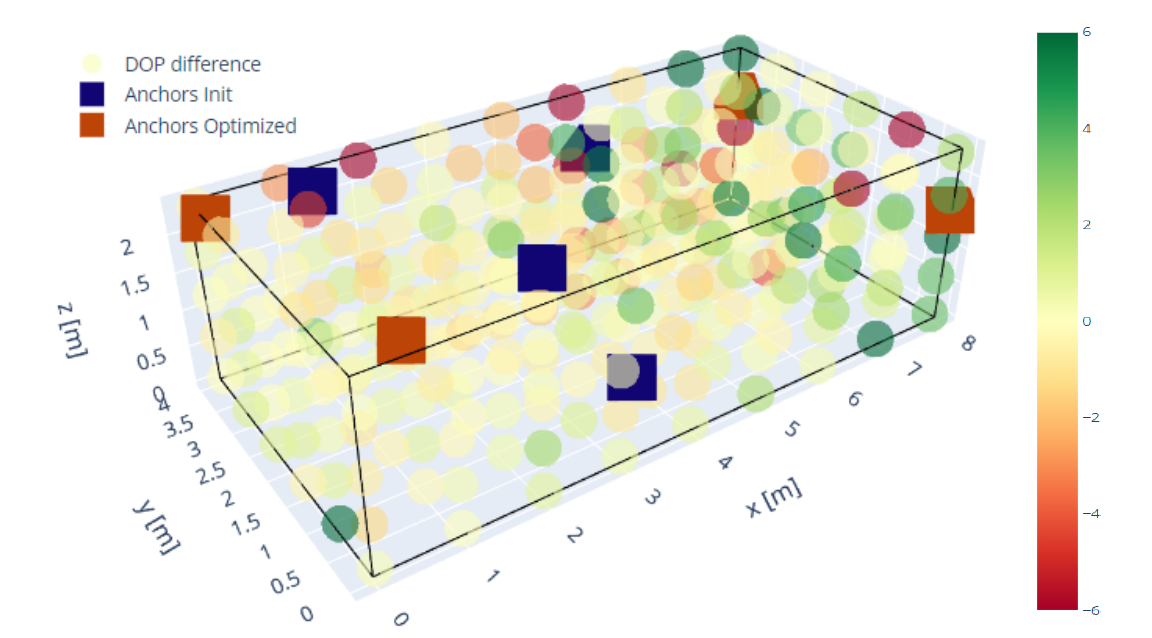}%
    \caption{\gls{dop} difference between the initialized and optimized 4 anchors scenario.}%
    \label{fig:dop_comp_4}
\end{figure}

The accompanying \gls{cdf} curves, shown in Fig.~\ref{fig:4anchCDFcurves}, indicate the relationship between x, y and z errors, in addition to the Euclidean distance error, between the optimized and non-optimized setup. 
During the optimization process, the accuracy along the x-direction drops slightly, when looking at the P95 value, to noticeably improve both the y and z accuracy and thus overall Euclidean distance accuracy.
\setlength{\belowcaptionskip}{-10pt}
\setlength{\belowcaptionskip}{0pt}
The evolution of accuracy improvement over \gls{pso} iterations can also be shown by the difference in obtained \gls{dop} values. 
The \gls{dop} provides an indication of how well the anchor geometry can determine a mobile node's position. 
After performing $R=500$ realizations of the specific initialization and optimal solution scenario, the standard deviation for the Euclidean distance $\sigma$ and ranging $\sigma_{\mathrm{r}}$ at each position can be determined. 
$\sigma_\mathrm{r}$ at a certain mobile node position in this case is the average of the standard deviations $\sigma_{\mathrm{r},j}$ obtained from each of the 4 anchors. 
The \gls{dop} for each position is defined as $\mathrm{DOP}^p=\sigma^p/\sigma_\mathrm{r}^p$. 
Fig~\ref{fig:dop_comp_4} shows the \gls{dop} improvement between the initialized and optimal scenario at each position as $\mathrm{DOP}_\text{init}^p-\mathrm{DOP}_\text{opt}^p$. 
Concluded from this figure, the \gls{dop} across all mobile node positions improves at most locations and deteriorates at few. 
On average, the \gls{dop} over all positions for these scenarios improved from $2.52$ to $1.74$. 

\setlength{\belowcaptionskip}{-10pt}
\setlength{\belowcaptionskip}{0pt}

\setlength{\belowcaptionskip}{-1pt}
\setcounter{figure}{5}
\begin{figure*}[!b] 
    \centering
    \subfloat[$t=1$, initial]{\includegraphics[width=0.66\columnwidth]{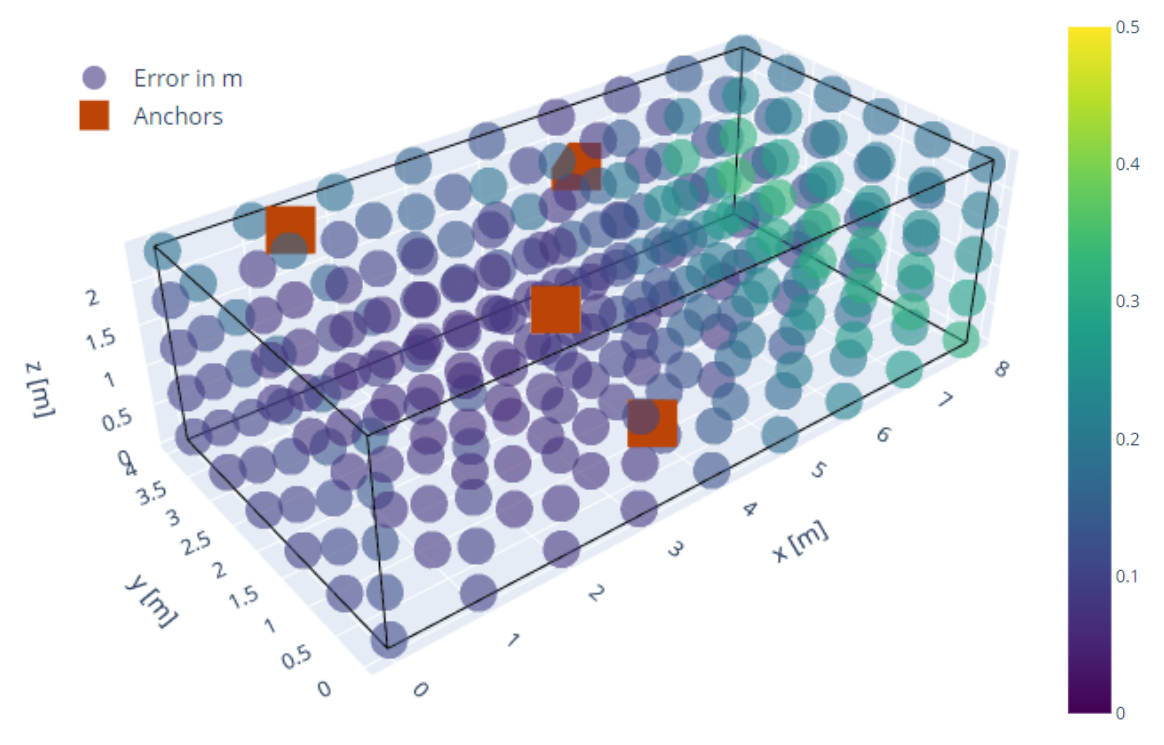}\vspace{-3mm}}\hfill%
    \subfloat[$t=20$]{\includegraphics[width=0.66\columnwidth]{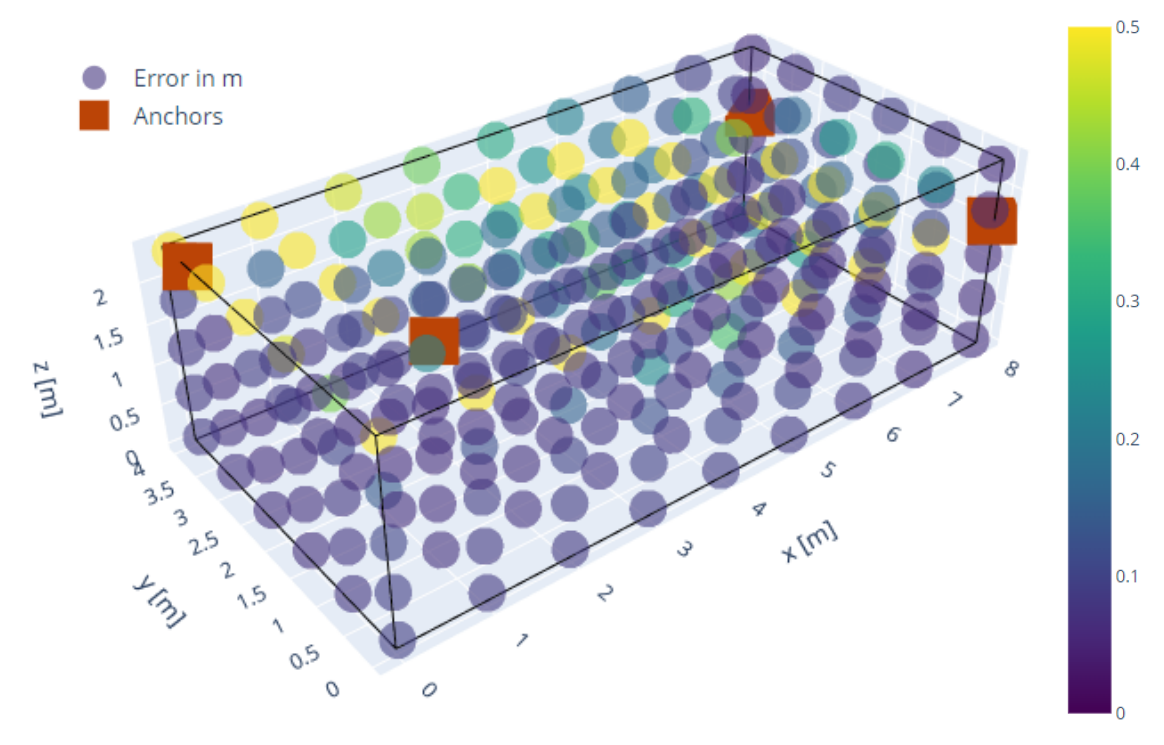}\vspace{-3mm}}\hfill%
    \subfloat[$t=40$, final]{\includegraphics[width=0.66\columnwidth]{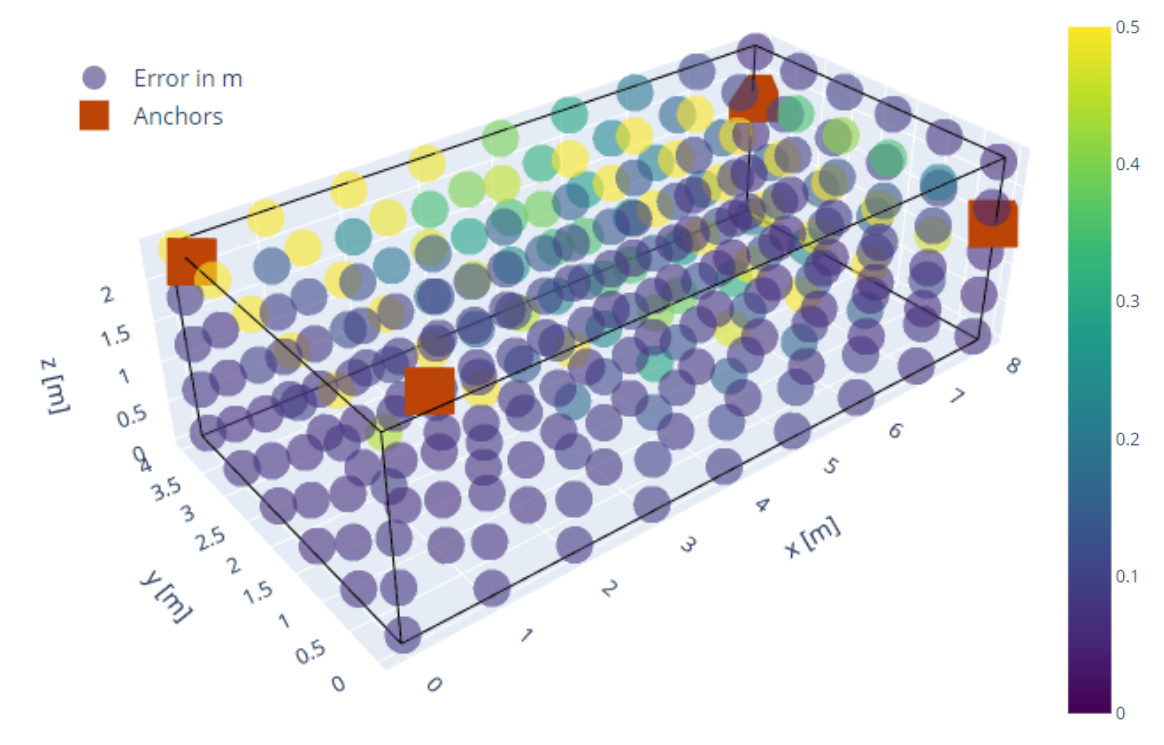}\vspace{-3mm}}\hfill%

    \caption{Evolution of the \gls{peb} based on the \gls{crlb} for the \gls{pso} algorithm iterations $t=\{1,20,40\}$ for $4$ anchors. } 
    \label{fig:3d_iterations}
\end{figure*}
\setlength{\belowcaptionskip}{0pt}

\setcounter{figure}{4}
\setlength{\belowcaptionskip}{-20pt}
\begin{figure}[!t]
\centering
    \centering
    \includegraphics[width=0.95\linewidth]{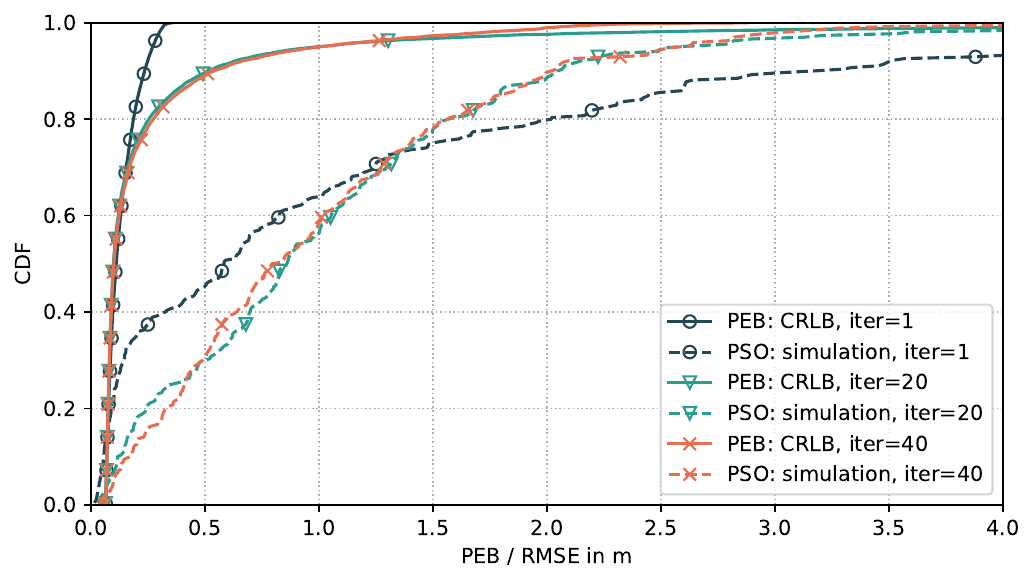}%
    \vspace{-2mm}
    \caption{Comparison between \glspl{cdf} of the \gls{rmse} and the \gls{peb} (via \gls{crlb}, neglecting outliers) obtained for the \gls{pso} iterations $t=\{1,20,40\}$ for $4$ anchors.} 
    \label{fig:cdf_iterations}
\end{figure}
\setlength{\belowcaptionskip}{0pt}

To investigate the performance of the \gls{pso} algorithm, the \gls{crlb} described in Section~\ref{sec:crlb} is compared to the \gls{rmse} of the position estimates for $R=500$ realizations.
For the numerical evaluation of the \gls{crlb}, the distance estimates were used to obtain the necessary range standard deviation for each anchor, which were in the range of $\sigma_\text{r} \approx 0.03-0.045~\mathrm{m}$. 
It should be noted that after removing outliers occurring at some positions, the range measurements for each anchor were reasonably well approximated as Gaussian distributed. 
Fig.~\ref{fig:cdf_iterations} shows the \glspl{cdf} of the \gls{peb} and the \gls{rmse} for iterations $t=\{1,20,40\}$ for $M=4$ anchors.
While the \gls{cdf} of the \gls{peb} shows a generally lower error over all candidate positions for the first iteration, the \gls{pso} algorithm results show an improvement, especially above the P95 value.
Most notably, for both the \gls{peb} as well as the \gls{rmse} curves, the only change from iteration $20$ to $40$ is the reduction of outliers, i.e., the maximum value of $1$ is reached at lower error values.
Comparing the first iteration, the \gls{rmse} initially follows the \gls{peb} until deviating, which can be explained by the fact that even though the range estimates are approximately Gaussian, outliers in the \gls{pso} simulation remain that are most likely attributable to overlapping propagation paths.
This overlap introduces outliers that are on the one hand not captured in the bounds, but also not easy to include. Therefore, the range measurement model used for the \gls{crlb} determination is too limited.
Nonetheless, the \gls{pso} simulation shows that the proposed optimization scheme is well suited to use for realistic scenarios, where a \gls{crlb} based analysis based on the given range measurement model could give over optimistic results and worse performance, especially for difficult measurement models.

Fig~\ref{fig:3d_iterations} shows the distribution of the \gls{peb} in 3D alongside the anchor positions per iteration.
While initially showing a ``favorable" anchor geometry for iteration $1$, the anchors converge to being positioned close to a plane, which results in a larger error perpendicular to that plane for agent positions on the plane. 
While non-optimal from a \gls{crlb} point of view, the \gls{cdf} comparison derived from the \gls{pso} simulations has shown that it is nonetheless well suited for that specific environment.



Overall it can be seen that the anchor placement layout can play a major role in the accuracy of an \gls{ips}. Another room or the same room with more or fewer mobile nodes may prefer a different optimal anchor placement layout, depending on the optimization parameter choice.

\subsection{Multiple Anchors}\label{sec:multiple_anchors}

Besides the anchor placement, the number of distributed anchors in the room has an influence on the accuracy and reliability of the \gls{ips}. 
To investigate this influence, situations are simulated with 4, 6, 8 and 10 anchors. 
An optimal anchor placement was first determined for each situation, as shown in Fig.~\ref{fig:multiplePlacement} using \gls{pso} to allow for a fair comparison between the best configurations. The standard deviation $\sigma$, mean $\mu$ and P95 values of the common Euclidean distance error are derived from simulation and summarised in Table~\ref{tab:var}. 

\setlength{\belowcaptionskip}{-10pt}
{\renewcommand{\arraystretch}{1.5} 
\begin{table}[h]

\centering
\begin{tabularx}{0.7\linewidth}{>{\hsize=0.4\hsize}L >{\hsize=0.15\hsize}c >{\hsize=0.15\hsize}c>{\hsize=0.15\hsize}c>{\hsize=0.15\hsize}c}
    \hline
    Anchors & 4 & 6 & 8 & 10 \\
    \hline
    $\sigma$ (m)& 0.803 & 0.691 & 0.587 &  0.563\\
    $\mu$ (m) & 0.995 & 0.913 & 0.797 & 0.773 \\
    P95 (m) &  2.629 & 2.272 & 1.986 & 1.876\\                          
    \hline
\end{tabularx}
\caption{Comparison between $\sigma$, $\mu$ and P95 values of the Euclidean distance errors for the 4, 6, 8 and 10 optimal anchor placement scenarios, derived from simulations.}
\label{tab:var}
\end{table}

}

\setlength{\belowcaptionskip}{0pt}

As expected, the standard deviation $\sigma$, mean $\mu$ and especially P95 values decrease as the number of anchors increases. The \gls{cdf} functions of the Euclidean distance error of the 3D position estimates alongside the \gls{peb} results are shown in Fig.~\ref{fig:cdf_multiple}. Both show an accuracy improvement when using multiple anchors. The achievable simulated positioning accuracy for this room shows minimal difference between 8 and 10 anchor nodes. Saturation of the accuracy values is observed for 8 anchors and no major improvement is expected with the addition of even more anchors. 
In more complex rooms with locations \gls{nlos} condition and possibly more reverberation, this saturation point is expected to be at a higher number.

While the use of multiple anchor nodes is desirable in terms of achievable accuracy, it is detrimental to the complexity, installation and update rate. 
The latter can be avoided by coding or by using \gls{fdm} instead of \gls{tdm} to avoid interference between anchors. 
If non-optimized anchor placements are used, the deployment of multiple anchors can ensure that a more optimal anchor combination, which is part of the non-optimized set, can be used to increase the precision at a specific location. 
However, this requires selection algorithms to choose the right set of anchors at a certain mobile node location. 
The anchor placement has a bigger impact on the precision and reliability than the amount of anchors used within an \gls{ips} employing the minimum amount of anchors necessary to perform unambiguous positioning. 
\setlength{\belowcaptionskip}{-10pt}
\setcounter{figure}{6}
\begin{figure}[t]
\centering
    \centering
    \includegraphics[width=1\linewidth]{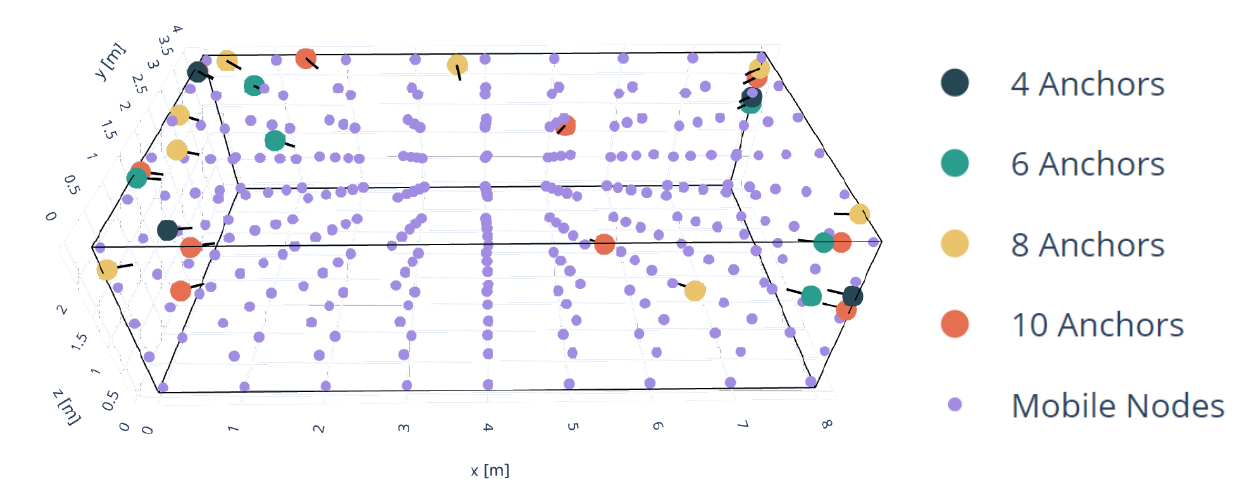}%
    \caption{Obtained anchor placement layout after \gls{pso}}%
    \label{fig:multiplePlacement}
\end{figure}

\setlength{\belowcaptionskip}{-15pt}
\begin{figure}[t]
\centering
    \centering
    \includegraphics[width=0.95\linewidth]{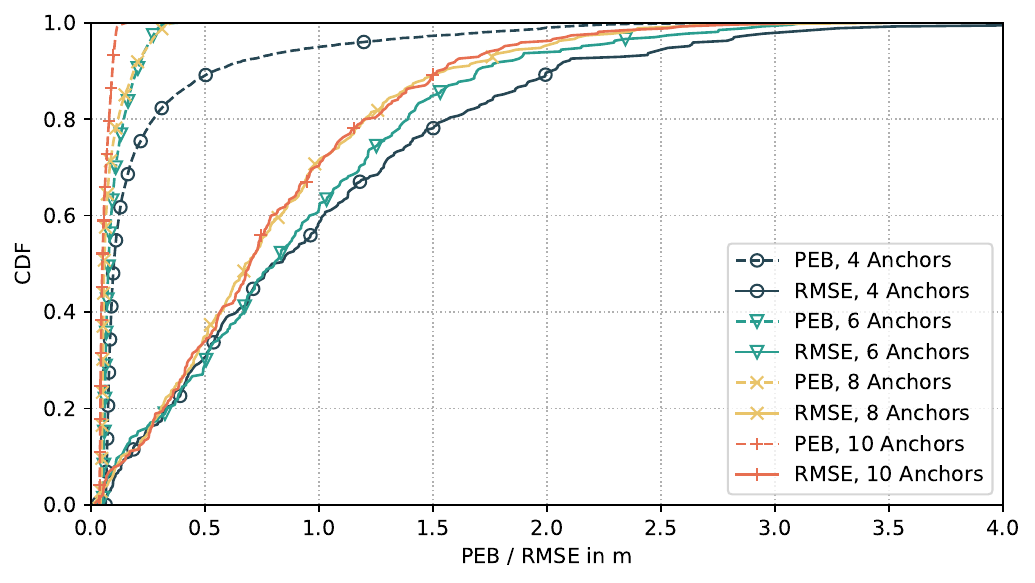}%
    \caption{Comparison between \glspl{cdf} of the simulated Euclidean 3D distance and \gls{peb} (via \gls{crlb}) obtained by the 4, 6, 8 and 10 optimal anchor placement scenarios.} 
    \label{fig:cdf_multiple}
\end{figure}
\setlength{\belowcaptionskip}{0pt}

\section{Conclusion} \label{sec:conclusion}
A \gls{pso} method to obtain an optimal anchor placement layout for a 3D \gls{ips} was presented. 
Based on simulations performed with this algorithm and a \gls{crlb} analysis, it can be concluded that a more optimal anchor placement layout for the \gls{ips} results in a desired increase in positioning accuracy and reliability within the entire room. There is even more to gain in having multiple anchors at distributed locations, taking into account an accuracy saturation at a certain number of anchors. An improvement of 30\% and 29\% on respectively the standard deviation and the P95 value is observed if 10 instead of 4 optimally placed anchors are applied. 
The improvements are expected to be even more extreme in advanced complex rooms with supplementary reverberation and \gls{nlos} effects. 
In addition, the cost function also plays an important role in developing the optimal anchor layout, with the actual choice of cost function being application dependent. 
Note that it is easy to adjust the cost function of the optimization method. 
The positioning system with optimal anchor placements will be validated via measurements in future work. 

\section*{Acknowledgment}
This work was supported by the Research Foundation-Flanders (FWO) under Grant G0D3819N.
This work has been partly supported by the European Union's Horizon 2020 research and innovation programme REINDEER project under grant agreement No. 101013425.

{\footnotesize \printbibliography}

@article{mdpi_bert,
	title        = {{Positioning energy-neutral devices: technological status and hybrid RF-acoustic experiments}},
	author       = {Cox, Bert and Buyle, Chesney and Delabie, Daan and De Strycker, Lieven and Van der Perre, Liesbet},
	year         = 2022,
	journal      = {Future Internet},
	volume       = 14,
	number       = 5,
	article-number = 156
}

@inproceedings{IPINBert,
	title        = {{Towards centimetre accurate and low-power, hybrid radio-acoustic 3D indoor positioning: an experimental journey.}},
	author       = {Cox, Bert and Buyle, Chesney and Van der Perre, Liesbet and De Strycker, Lieven},
	year         = 2021,
	booktitle    = {2021 IEEE 
Indoor Positioning and Indoor Navigation (IPIN)},
	organization = {CEUR Workshop}
}

@inproceedings{techtile-acoustic,
	title        = {Techtile: a Flexible Testbed for Distributed Acoustic Indoor Positioning and Sensing},
	author       = {Delabie, Daan and Cox, Bert and De Strycker, Lieven and Van der Perre, Liesbet},
	year         = 2022,
	booktitle    = {IEEE Sensors Applications Symposium (SAS)},
	volume       = {},
	number       = {},
	pages        = {}
}

@inproceedings{CRLBfirst,
	title        = {Incremental anchor layout for indoor positioning},
	author       = {Qiao, Tianzhu and Liu, Huaping},
	year         = 2017,
	booktitle    = {2017 IEEE International Conference on Communications (ICC)},
	volume       = {},
	number       = {},
	pages        = {1--6}
}

@inproceedings{SimulationPaper,
	title        = {An Acoustic Simulation Framework to Support Indoor Positioning and Data Driven Signal Processing Assessments},
	author       = {Daan Delabie and Chesney Buyle and Bert Cox and Liesbet Van der Perre and Lieven De Strycker},
	year         = 2023,
        booktitle    = {2023 IEEE European Signal Processing Conference (EUSIPCO)},
        eprint={2305.02715},
      archivePrefix={arXiv},
}

@article{PSOoverview,
	title        = {Particle Swarm Optimization Algorithm and Its Applications: A Systematic Review},
	author       = {Gad, Ahmed},
	year         = 2022,
	month        = {04},
	journal      = {Archives of Computational Methods in Engineering},
	volume       = 29,
	pages        = {}
}

@article{PSO2,
	title        = {Particle Swarm Optimisation: A Historical Review Up to the Current Developments},
	author       = {Freitas, Diogo and Lopes, Luiz Guerreiro and Morgado-Dias, Fernando},
	year         = 2020,
	journal      = {Entropy},
	volume       = 22,
	number       = 3,
	article-number = 362
}

@article{GA_example,
	title        = {Comparison of Objectives in Multiobjective Optimization of Ultrasonic Positioning Anchor Placement},
	author       = {Haigh, Sebastian and Kulon, Janusz and Partlow, Adam and Gibson, Colin},
	year         = 2022,
	journal      = {IEEE Transactions on Instrumentation and Measurement},
	volume       = 71,
	number       = {},
	pages        = {1--12}
}

@inproceedings{OnOptimalAnchorPlacement,
	title        = {On optimal anchor placement for efficient area-based localization in wireless networks},
	author       = {Lasla, Noureddine and Younis, Mohamed and Ouadjaout, Abdelraouf and Badache, Nadjib},
	year         = 2015,
	booktitle    = {2015 IEEE International Conference on Communications (ICC)},
	volume       = {},
	number       = {},
	pages        = {3257--3262}
}

@article{practical_amount_drone,
	title        = {Sound Source Localization in Wide-Range Outdoor Environment Using Distributed Sensor Network},
	author       = {Faraji, Mohammad Mahdi and Shouraki, Saeed Bagheri and Iranmehr, Ensieh and Linares-Barranco, Bernabe},
	year         = 2020,
	journal      = {IEEE Sensors Journal},
	volume       = 20,
	number       = 4,
	pages        = {2234--2246}
}

@article{acoustic_self_loc,
	title        = {Acoustic self-localization in a distributed sensor network},
	author       = {Frampton, K.D.},
	year         = 2006,
	journal      = {IEEE Sensors Journal},
	volume       = 6,
	number       = 1,
	pages        = {166--172}
}

@article{OnOptimal2,
	title        = {On Optimal Anchor Placement for Area-based Localization in Wireless Sensor Networks},
	author       = {Cheriet, Abdelhakim and Bachir, Abdelmalik and Lasla, Noureddine and Abdallah, Mohamed},
	year         = 2021,
	month        = {04},
	journal      = {IET Wireless Sensor Systems},
	volume       = 11,
	pages        = {}
}

@inproceedings{optimalAhmad,
	title        = {Optimal anchors placement strategy for super accurate nodes localization in anisotropic wireless sensor networks},
	author       = {El Assaf, Ahmad and Zaidi, Slim and Affes, Sofiène and Kandil, Nahi},
	year         = 2016,
	booktitle    = {2016 Intl. Wireless Communications and Mobile Computing Conference (IWCMC)},
	volume       = {},
	number       = {},
	pages        = {25--30}
}

@article{PSOanchorsExample,
	title        = {An Optimal Anchor Placement Method for Localization in Large-Scale Wireless Sensor Networks},
	author       = {Tugrul \c{C}avdar and Faruk Baturalp Günay and Nader Ebrahimpour and Muhammet Talha Kakız},
	year         = 2022,
	journal      = {Intelligent Automation \& Soft Computing},
	volume       = 31,
	number       = 2,
	pages        = {1197--1222}
}

@book{kay1993,
	title        = {Fundamentals of Statistical Signal Processing, Vol. I: Estimation Theory},
	author       = {Kay, Steven M},
	year         = 1993,
	publisher    = {Signal Processing. Upper Saddle River, NJ: Prentice Hall}
}

@article{GaussianBert,
	title        = {{High precision hybrid RF and ultrasonic chirp-based ranging for low-power IoT nodes}},
	author       = {Cox, Bert and Van der Perre, Liesbet and Wielandt, Stijn and Ottoy, Geoffrey and De Strycker, Lieven},
	year         = {2020},
	journal      = {Eurasip Journal On Wireless Communications And Networking},
	publisher    = {Springer},
	number       = 1,
	pages        = {1--24},
}

@article{GuvencCST2009,
	title        = {{A survey on TOA based wireless localization and NLOS mitigation techniques}},
	author       = {Guvenc, Ismail and Chong, Chia-Chin},
	year         = 2009,
	journal      = {IEEE Communications Surveys \& Tutorials},
	volume       = 11,
	number       = 3,
	pages        = {107--124}
}

@INPROCEEDINGS{Geometrical_ipin,
  author={Sharma, Ravi and Badarla, Venkataramana},
  booktitle={2018 IEEE International Conference on Advanced Networks and Telecommunications Systems (ANTS)}, 
  title={{Geometrical optimization of a novel beacon placement strategy for 3D indoor localization}}, 
  year={2018},
  volume={},
  number={},
  pages={1-6}
}

\end{document}